\documentclass[11pt,a4paper]{article}
\usepackage{jheppub} 
\usepackage{bm}
\usepackage{slashed}

\newcommand{\beq}{\begin{equation}} 
\newcommand{\eeq}{\end{equation}}
\newcommand{\beqa}{\begin{eqnarray}} 
\newcommand{\eeqa}{\end{eqnarray}} 
\newcommand{\non}{\nonumber}

\title{Parity odd equilibrium partition function in $2+1$ dimensions}

\author[a]{Juan L. Ma\~nes,}
\author[b]{Manuel Valle}

\affiliation[a]{Departamento de F\'\i sica de la Materia Condensada, 
Universidad del Pa\'is Vasco UPV/EHU, \\
Apartado 644,  48080 Bilbao, Spain}
\affiliation[b]{Departamento de F\'\i sica Te\'orica, 
Universidad del Pa\'is Vasco UPV/EHU, \\
Apartado 644,  48080 Bilbao, Spain}

\abstract{

We use Schwinger's proper time method to compute the parity odd contributions  to the $U(1)$ current  and energy-momentum tensor of  an ideal gas of  fermions in \hbox{$2+1$} dimensions in the presence of static gauge and gravitational backgrounds. From these results  the equilibrium partition function at first order in the derivative expansion is explicitly obtained by integration. The form of the computed partition function is consistent with general arguments based on Kaluza-Klein and gauge invariance.

}

\begin{document}
\maketitle
\flushbottom

\section{Introduction}

Apart from the direct thermodynamic description, 
the equilibrium partition function  as a functional of  stationary gauge and gravitational fields 
plays an important role in  the realm of fluid dynamics. 
In particular, the structure of the first few  terms  of an expansion in derivatives of these fields has an interplay 
with the non-dissipative part of the constitutive relations of hydrodynamics, which connect expectation values of microscopic currents with 
macroscopic fields such as  temperature, chemical potential and fluid velocities. 
This  has led to significant progress in understanding 
some  parity-odd 
susceptibilities 
related to anomalous conservation equations  
of field theories with anomalies~\cite{Son:2009tf, Neiman:2010zi, Bhattacharya:2011eea, Landsteiner:2011cp, Bhattacharya:2011tra, 
Loganayagam:2011mu, Jensen:2011xb, Loganayagam:2012pz, Jain:2012rh, Valle:2012em, Jensen:2012kj}, 
without the need to resort to entropy production arguments~\cite{Dubovsky:2011sk, Dubovsky:2011sj,  
Banerjee:2012iz, Jensen:2012jh}.

The general form of the thermal equilibrium partition function that includes 
first order terms in the derivative expansion of the background fields has been discussed  previously  (see e. g.~\cite{Banerjee:2012iz}). 
For an arbitrary time-independent background given by the 
 line element and $U(1)$ gauge connection 
 \beq\label{ds2}
\begin{split}
ds^2&=-e^{2\sigma(\bm{x})} (dt+a_i(\bm{  x}) dx^i)^2 +g_{ij}(\bm{x}) dx^idx^j, \qquad  i, j=1,\ldots p, \\ 
A_\mu&=(A_0(\bm{x}), \bm{A}(\bm{x})) \,, 
\end{split}
\eeq
the partition function in one and three spatial dimensions depends only  on a few constant coefficients, which appear 
in the  anomalous conservation equations.  
In two spatial dimensions the form of the corresponding partition function is less restricted. 
In this case  there are not anomalous conservation laws, but the  mass  of the Dirac field  acts as a source of parity violation, and 
it turns out that  the partition function may depend on arbitrary functions  of the mass and certain combinations of  the background fields. 
Arguments based on Kaluza-Klein and gauge invariance~\cite{Banerjee:2012iz} constrain   
the form of the most general parity-odd partition function  at first derivative order in $2+1$ dimensions  
 to be 
\beq\label{part1}
\mathcal{W}=\frac{1}{2}\int d^2x \left({ \alpha}(\sigma,A_0)\epsilon^{ij}\partial_i \tilde A_j+T_0\beta(\sigma,A_0)\epsilon^{ij}\partial_i a_j\right) , 
\eeq
where $ \tilde A_i=A_i-A_0 a_i$,  $T_0$ is the equilibrium temperature and the  functions  ${\alpha}(\sigma,A_0)$ and  ${\beta}(\sigma,A_0)$ are 
specific to the system considered.

The main purpose of this paper is precisely the computation of the parity violating partition function 
for a noninteracting massive Dirac fermion in 2+1 
dimensions  at  first order in the derivatives of the background fields. 
In order to do this  we first obtain the parity-odd contributions to the $U(1)$ currents and energy-momentum tensor in the non-uniform static background defined by the gauge field and the metric~\eqref{ds2} and  identify our results with  the  variational formulas from $\mathcal{W}$. Integrating the corresponding equations we are able to completely determine the functions $\alpha(\sigma,A_0)$ and $\beta(\sigma,A_0)$. 
As far as we know, the  piece in~\eqref{part1} proportional to the curl of the vector field $\bm{a}$ had never been   computed  before and this is one of our main results. 

We find that the functions $\alpha$ and $\beta$ thus determined vanish for massless  fermions. This, however, does not imply the absence of parity violating effects in this limit. The theory has to be regularized, and 
a convenient gauge-invariant regularization procedure in two spatial dimensions is  the Pauli-Villars method, 
which uses massive fermions as regulator fields. From a study of the asymptotic behavior of the partition function for large fermion masses we conclude that,  at first order in the derivative expansion,  there are   parity-odd contributions to the currents of massless fermions ---which produce the nonzero Hall currents from the parity anomaly~\cite{Niemi:1983rq, Redlich:1983kn, Redlich:1983dv,  AlvarezGaume:1984nf}--- but not to their energy-momentum tensor. This is consistent with the fact that the purely gravitational  contribution at zero temperature in 2+1 dimensions
is described by the gravitational Chern-Simons action,  which gives rise to 
the   third (derivative) order Cotton tensor~\cite{Deser:1981wh,AlvarezGaume:1984nf}.

The contents of this paper is organized  as follows. 
In section~\ref{general} we express the components of the currents and energy-momentum tensor in terms of  the coincidence limit 
of the thermal Green function  $\mathcal{G}( x,  x',\omega_n)$ and its first derivatives. The derivative expansion for the thermal Green function is computed in section~\ref{derexp} using Schwinger's proper time method, and from the Green function  the parity-odd contributions to the currents and energy-momentum tensor are derived in section~\ref
{compcurr}. These results are used to determine the partition function in  section~\ref{partition}, where we also discuss
the small and large fermion mass limits and the effects of a Pauli-Villars regulator. Our conclusions and possible extensions of this work are presented in section~\ref{discuss}.

 \section{General expressions for the $U(1)$ current and energy-momentum tensor}
\label{general} 
In this section we  obtain general formulas for the current and energy-momentum tensor in a static background and express their components in terms of  the coincidence limit 
of the thermal Green function   and its first derivatives. 
The action for a charged spin-$1/2$ fermion in a curved background is given by $S=\int d^3 x\sqrt{-g}\mathcal L$, where 
\begin{equation}
\mathcal{L}=-i\bar\psi\underline\gamma^\mu\nabla_\mu\psi +im\bar\psi\psi
\end{equation}
and the space-time dependent  matrices $\underline\gamma^\mu(x)$    are related to the constant Minkoswki   matrices by $\underline\gamma^\mu(x)=e^\mu_a(x) \gamma^a$, where $e_a^\nu$ is  the vielbein. They satisfy   $\{	\underline\gamma^\mu(x),	\underline\gamma^\nu(x)\}=2g^{\mu\nu}(x)$ and $\{\gamma^a,\gamma^b\}=2\eta^{ab}$, with $\eta^{a b} = \text{diag}(-1,1,1)$ in $2+1$ dimensions.
We choose a representation where the constant Minkowski matrices are given in terms  of the Pauli matrices by
 \beq
\gamma^0=-i\sigma_3\;\;\; , \;\;\;  \gamma^1=\sigma_2 \;\;\; , \;\;\; \gamma^2=-\sigma_1\ .
\eeq
 Note that  $\bar \psi$ is related to $\psi^\dagger$ by  $\bar \psi=\psi^\dagger\gamma^0=-i \psi^\dagger \sigma_3 $.

The covariant derivative of the Dirac field $\psi$ is given by~\cite{Birrell:1982ix,Freedman:2012zz}
\beq
\nabla_\mu \psi = (\partial_\mu + \frac{1}{4} \omega_\mu^{\;\; a  b} \gamma_{a b} - i A_\mu ) \psi , 
           \qquad \gamma_{a b} = \frac{1}{2} [\gamma_a, \gamma_b]\ ,
\eeq
where, in the absence of torsion, the spin connection is related to the vielbein $e_a^\nu$ 
 by
\beq
\omega_\mu^{\;\; a  b} = -e^{b \nu} (\partial_\mu e_\nu^{a} - \Gamma_{\mu \nu}^{\sigma} e_\sigma^a)\ .
\eeq
Here $\Gamma_{\mu \nu}^{\sigma}$ are the   Christoffel symbols obtained from the metric.
For the rest of the paper, we will consider a static metric of the  form
\beq\label{metric}
ds^2=-e^{2\sigma(\bm  x)} (dt+a_i (\bm  x) dx^i)^2 +\delta_{ij}dx^idx^j\ ,\;\;\; i,j=1,2
\eeq
 and  assume that the gauge fields are time-independent as well. Although the most general static metric would have $g_{ij}\neq \delta_{ij}$, the one considered here can be used to obtain all the parity-odd contributions to the $U(1)$ current and energy-momentum tensor. This form of the metric  is preserved by  redefinitions of time of the form  
 \beq\label{kk}
 t'=t+\phi(\bm  x)
 \eeq
  with the Kaluza-Klein (KK) field $a_i$ transforming like a connection
 \beq
 a'_i=a_i-\partial_t\phi \ .
 \eeq
As pointed out  in~\cite{Banerjee:2012iz}, one can easily check that tensors with upper spatial indices and lower temporal indices are automatically invariant under the KK transformation~\eqref{kk}. Combinations of the form $V_i-a_i V_0$ are also KK-invariant.

Using the variational definition of the $U(1)$ current 
 \beq
J_\mu=\frac{g_{\mu\nu}}{\sqrt{-g}}\frac{\delta S}{\delta A_\mu}
\eeq
yields the following formulas for the KK-invariant components

\beq
J_0=-e^\sigma\psi^\dagger\psi\;\;\; ,\;\;\; J^i=\psi^\dagger\sigma_i\psi\;\;\; ,\;\;\; i=1,2\ .
\eeq

Similarly, using the standard formula for the energy-momentum tensor~\cite{Birrell:1982ix,Freedman:2012zz}
\beq
T_{\mu \nu} =  \frac{i}{4} \bar{ \psi} \left[ \gamma_\mu \overrightarrow{\nabla}_\nu  - 
       \overleftarrow{\nabla}_\nu \gamma_\mu + ( \mu \leftrightarrow \nu) 
      \right ] \psi 
\eeq
with
\beq
\bar{ \psi}\, \overleftarrow{\nabla}_\mu = \bar{ \psi}( \overleftarrow\partial_\mu  - \frac{1}{4} \omega_\mu^{\;\; a  b} \gamma_{a b} + i A_\mu )\ , 
\eeq
 the following expressions are obtained 

\beq\label{class}
\begin{split}
T_{00} &= \frac{i}{2} e^{\sigma} (\psi^\dagger \partial_t \psi -  \partial_t \psi^\dagger \, \psi) + e^{\sigma}  A_0 \, \psi^\dagger  \psi 
                  -\frac{1}{4} e^{3 \sigma} \epsilon^{i j} \partial_i a_j \,  \psi^\dagger \sigma_3 \psi\ , \\ 
T_{0}^i &= \frac{i}{4} e^{\sigma} (\psi^\dagger \partial_i  \psi -  \partial_i \psi^\dagger \, \psi) + \frac{1}{2} e^{\sigma}  A_i \, \psi^\dagger  \psi +\frac{1}{4} e^{\sigma} \epsilon^{i j} \partial_j \sigma \,  \psi^\dagger \sigma_3 \psi \\ 
                                    & \quad - \frac{i}{4}  (\psi^\dagger \sigma_i \partial_t  \psi -  \partial_t \psi^\dagger \sigma_i \psi) 
                        - \frac{1}{2}   A_0 \,  \psi^\dagger \sigma_i   \psi \\ 
                  & \quad - \frac{i}{4} e^\sigma a_i  (\psi^\dagger  \partial_t  \psi -  \partial_t \psi^\dagger \, \psi) 
                        - \frac{1}{2} e^\sigma a_i    A_0 \,  \psi^\dagger   \psi\ .
                                             \end{split}
\eeq 
On general grounds~\cite{Banerjee:2012iz}  the KK-invariant components  $T^{ij}$  can have no parity-odd contributions at first derivative order and will not be considered   here. See, however, our comments at the end of section~\ref{discuss}.

In the next section we will compute a derivative expansion for the thermal Green function  $\mathcal{G}( \bm  x,  \bm  x',\omega_n)$ defined by~\cite{lebellac}
\beq\label{green}
\langle T\psi(-i\tau,  \bm  x)\psi^\dagger(0, \bm  x') \rangle_\beta=T_0\sum_n e^{-i\omega_n\tau} \mathcal{G}( \bm  x,  \bm  x',\omega_n)\ ,
\eeq
where the sum runs over fermionic Matsubara frequencies $\omega_n=\frac{2\pi}{\beta} (n+\frac{1}{2})$ and  $\beta=1/T_0$. In terms of this Green functions the currents and energy-momentum tensor are given by
\beq\label{curr}
\begin{split}
 J^i&=T_0\sum_n\left[ \mathrm{tr}\, \sigma_i\,\mathcal{G}(\bm  x, \bm  x, \omega_n)\right]\ ,\\
 J_0&=T_0\sum_n\left[-e^\sigma \mathrm{tr}\, \mathcal{G}(\bm  x, \bm  x, \omega_n)\right]
\end{split}
\eeq
and
\beq\label{tmunu}
\begin{split}
T_{00} &= T_0 \sum_n \left[ e^\sigma (i \omega_n + A_0)\,  \text{tr}\, \mathcal{G}(\bm  x, \bm  x, \omega_n) - 
    \frac{1}{4} e^{3 \sigma}  \epsilon^{i j} \partial_i a_j\,   \text{tr}\, \sigma_3\, \mathcal{G}(\bm  x,\bm  x, \omega_n)  \right] , \\ 
T_{0}^i &= T_0 \sum_n \left[ \frac{i}{4} e^\sigma   \text{tr}\left( \frac{\partial}{\partial x^i}\mathcal{G}(\bm  x, \bm  x', \omega_n)  - 
         \frac{\partial}{\partial x^{\prime i}}\mathcal{G}(\bm  x, vx', \omega_n) \right)  \right] \biggl\lvert_{\bm  x' = \bm  x}  \\ 
         &\quad +  T_0 \sum_n\left[  \frac{1}{2} e^\sigma A_i\,  \text{tr}\, \mathcal{G}(\bm  x, \bm  x, \omega_n)+ \frac{1}{4} e^{\sigma} \epsilon^{i j} \partial_j \sigma \,  \text{tr}\, \sigma_3 \mathcal{G}(\bm  x,\bm  x, \omega_n)\right]  \\ 
                 &\quad + T_0 \sum_n \left[ -\frac{1}{2} (i \omega_n + A_0)\,  \text{tr}\, \sigma_i \mathcal{G}(\bm  x, \bm  x, \omega_n)\right]  \\ 
         &\quad + T_0 \sum_n \left[ -\frac{1}{2} e^\sigma a_i (i \omega_n + A_0)\,  \text{tr}\, \mathcal{G}(\bm  x, \bm  x, \omega_n)\right]  . 
\end{split}
\eeq

\section{Derivative expansion of the Green function}
\label{derexp}
In this section we compute the two-point Green function for an ideal gas of fermions  at finite density and temperature in the static $U(1)$ and gravitational background considered in the previous section. As a first step, we rewrite the action as
\beq
S=-\int d^3 x\, \psi^\dagger \sqrt{-g} \,\gamma^0 \underline\gamma^0 \left[i \partial_t-\mathcal{H}\right]\psi\,,
\eeq 
where the hamiltonian density operator $\mathcal{H}$ is given by~\cite{Huang:2008kh}
\beq\label{ham}
\mathcal{H}=-i(\frac{1}{4} \omega_0^{\;\; a  b}-i A_0) -\frac{im}{g^{00}}\, \underline\gamma^0-\frac{i}{g^{00}}\, \underline\gamma^0  \underline\gamma^k\,\nabla_k\;\;\; ,\;\;\; k=1,2\,.
\eeq
Next we rotate to euclidean time  $t\to-i\tau$ and consider the thermal Green function  $\mathcal{G}( \bm  x,  \bm  x',\omega_n)$ defined in Eq.~\eqref{green}.
This Green function satisfies the following differential equation
\beq\label{eqgreen}
-\sqrt{-g}\gamma^0 \underline\gamma^0(i\omega_n-\mathcal{H})\mathcal{G}( \bm  x,  \bm  x',\omega_n)= \delta( \bm  x- \bm  x')\,.
\eeq
For the background metric~\eqref{metric}, the prefactor  takes the simple form
\beq
-\sqrt{-g}\gamma^0 \underline\gamma^0=\mathbf{1}-e^\sigma \boldsymbol a\cdot \boldsymbol\sigma\,,
\eeq
where $\mathbf{1}$ is the $2\times 2$ unit matrix. By construction, the hamiltonian~\eqref{ham} depends only on the background fields and their first derivatives. After some lengthy algebra we obtain 
\beq
-\sqrt{-g}\gamma^0 \underline\gamma^0\,\mathcal{H}=H_0+H_1\,,
\eeq
where
\beqa
H_0&=&-A_0\mathbf{1}+e^\sigma m\sigma_3- e^\sigma\boldsymbol\sigma\cdot(i\boldsymbol\partial+\boldsymbol A-A_0\,\boldsymbol a) \\
H_1&=&-\frac{e^{2\sigma}}{4}\epsilon^{ij}\partial_i a_j\,\sigma_3-\frac{i}{2}e^\sigma\boldsymbol\sigma\cdot\boldsymbol\partial\,\sigma\,.
\eeqa
Then Eq.~\eqref{eqgreen} can be written
\beq\label{eqgreen1}
\left[ (\mathbf{1}-e^\sigma\, \boldsymbol a\cdot \boldsymbol\sigma)i\omega_n-H( x)\right]\mathcal{G}( \bm  x,  \bm  x',\omega_n)=\delta( \bm  x- \bm  x')\,,
\eeq
where $H(\bm   x)=H_0( \bm  x)+H_1( \bm  x)$. For arbitrary non-uniform background fields  this equation can not be solved in closed form, but we may consider a derivative expansion
\beq
\mathcal{G}\sim \mathcal{G}_0+\mathcal{G}_1+\ldots\,,
\eeq
where the subscript indicates the order in  derivatives of the background fields. In particular, $\mathcal{G}_0$ is the Green function for a  constant background specified by the values of $A_0, \sigma, \boldsymbol a$ and  $\boldsymbol A$ at some fixed reference point.


\subsection{The Green function at leading order}
The Green function at leading order is obtained by neglecting   the derivatives of the background fields. Then Eq.~\eqref{eqgreen1} simplifies to
\beq\label{eqgreen0}
\left[ (\mathbf{1}-e^{\sigma(z)}\, \boldsymbol a(\bm  z) \cdot \boldsymbol\sigma)i\omega_n-{H}_0( \bm  z)\right]\mathcal{G}_0( \bm  x,  \bm  x',\omega_n)=\delta( \bm  x- \bm  x')\,,
\eeq 
where the background fields are evaluated at the reference point $\bm  z$ but the derivatives in $H(\bm  z)$ act on $\bm  x$
\beq\label{h0}
{H}_0( \bm  z)=-A_0(\bm  z)\mathbf{1}+e^{\sigma(\bm  z)} m\sigma_3- e^{\sigma(\bm  z)}\sigma_i\left(i\frac{\partial}{\partial x^i}+ A_i(\bm  z)-A_0(\bm  z)\, a_i(\bm  z)\right).
\eeq
Taking the space Fourier transform   gives
 \beq\label{G0F}
\mathcal{G}_0(\bm  p,\omega_n)=\frac{(A_0+i\omega_n)\mathbf{1}+m e^\sigma\sigma_3+e^\sigma\boldsymbol \sigma\cdot\boldsymbol{\tilde p}}{(A_0+i\omega_n)^2 -e^{2\sigma}(m^2+\tilde {\bm p}^2)}\,,
\eeq
where 
\beq\label{shift}
\tilde p_i\equiv p_i- A_i+(A_0+i \omega_n) \,a_i\,.
\eeq
Note the occurrence of the KK-invariant combination $A_i-A_0 \,a_i$. As Eq.~\eqref{shift} implies the following relation in position space
\beq
\mathcal{G}_0( \bm  x,  \bm  x',\omega_n)=e^{i(\boldsymbol  A-(A_0+i\omega_n)\,\boldsymbol a)\cdot(\boldsymbol x-\boldsymbol x')}\Bigl.\,\mathcal{G}_0( \bm  x,  \bm  x',\omega_n)\Bigr|_{\boldsymbol A=\boldsymbol a=0}\;,
\eeq
we may restrict ourselves to the case $\boldsymbol A=\boldsymbol a=0$, adding the phase at the end of the computation. 
Given the structure of~\eqref{G0F}, it  is convenient to consider first the scalar Green function with Fourier transform
\beq
\Delta_0(\bm  p,\omega_n)=\frac{1}{(A_0+i\omega_n)^2 -e^{2\sigma}(m^2+\bm  p^2)}\,,
\eeq
which in position space is given by 
\beq
\Delta_0( \bm  x,  \bm  x',\omega_n)=-\frac{1}{2\pi}e^{-2\sigma}K_0\left(|\bm  x-\bm  x'|\sqrt{m^2-e^{-2\sigma}(A_0+i\omega_n)^2}\right)\,,
\eeq
where $K_0$ is the modified Bessel function of the second kind. Then the fermionic Green function at leading order is given by
\beq\label{rel}
\mathcal{G}_0( \bm  x,  \bm  x',\omega_n)=e^{i(\boldsymbol  A-(A_0+i\omega_n)\,\boldsymbol a)\cdot(\boldsymbol x-\boldsymbol x')}\,\mathcal{D}_L(x)\Delta_0( \bm  x,  \bm  x',\omega_n)\,,
\eeq
where $\mathcal{D}_L(\bm  x)$ is the Fourier transform of the numerator in~\eqref{G0F}
\beq
\mathcal{D}_L(\bm  x)=(A_0+i\omega_n)\mathbf{1}+m\, e^\sigma\sigma_3-ie^\sigma\boldsymbol \sigma\cdot\boldsymbol\partial\,.
\eeq
This finally yields the following explicit expression for the Green function ay leading order
\beq\label{green0}
\begin{split}
\mathcal{G}_0( \bm  x,  \bm  x',\omega_n)&=-\frac{e^{i(\boldsymbol A-(A_0+i\omega_n)\boldsymbol a)\cdot(\boldsymbol x-\boldsymbol x')-2\sigma}}{2\pi}\left[((A_0+i\omega_n)\mathbf{1}+m\, e^\sigma\sigma_3)K_0(b |\bm  x-\bm  x'|)\right.\\
&\left. +i( \boldsymbol {\hat x} \cdot\!\boldsymbol \sigma) b K_1 (b| \bm  x-\bm  x'| )\right]\,,
\end{split}
\eeq
where 
\beq\label{b2}
b^2\equiv m^2-e^{-2\sigma}(A_0+i\omega_n)^2
\eeq
and we have used
\beq
\partial_i K_0(b| \bm  x|)=-\frac{ x_i}{|\bm   x|} b K_1(b| \bm  x|)\,.
\eeq
Remember that in~\eqref{green0} all the fields are evaluated at the reference point $\bm z$, which in general will be different from $\bm x$ and $\bm x'$. 

\subsection{The Green function at first derivative order}
Here we will use perturbation theory to compute $\mathcal{G}_1(i\omega_n,\bm  x,  \bm  x')$, i.e., the contribution to the Green function at first order in the derivatives of the background fields. In order to simplify the calculations  we will assume that, by a combination of  gauge and KK transformations, the fields $	\boldsymbol A$ and $	\boldsymbol a$ have been set to zero at a reference point $\bm  z$. As explained at the end of this section, the complete dependence  on $\boldsymbol A$ and $	\boldsymbol a$  of physical observables computed from  $\mathcal{G}_1(i\omega_n,\bm  x,  \bm  x')$ can be restored at the end of the calculation by invoking gauge and KK invariance. We  expand the fields around the reference point
\beq\label{ref}
\begin{split}
A_0(\bm  x)&\simeq A_0(\bm  z) +(x^i-z^i)\partial_iA_0(\bm  z)\\
\sigma(\bm  x)&\simeq \sigma(\bm  z) +(x^i-z^i)\partial_i\sigma(\bm  z)\\
A_j(\bm  x)&\simeq (x^i-z^i)\partial_iA_j(\bm  z)\\
a_j(\bm  x)&\simeq (x^i-z^i)\partial_ia_j(\bm  z)
\end{split}
\eeq
and, correspondingly, the hamiltonian
\beq\label{exph}
{H}(\bm  x)={H}_0(\bm  z)+\delta{H}(\bm  x)+\ldots\,,
\eeq
where $\delta{H}(\bm  x)$ is linear in the derivatives of the fields and is given by
\beq
\delta{H}(\bm  x)=(x^i-z^i)\left.\partial_i H_0\right|_z+H_1(\bm  z)\,.
\eeq
We must also expand the prefactor
\beq\label{expp}
-\sqrt{-g}\gamma^0 \underline\gamma^0(\bm  x)\simeq\mathbf{1}-e^{\sigma(\bm  z)}\delta \boldsymbol a(\bm  x)\cdot \boldsymbol\sigma\equiv\mathbf{1}-(x^j-z^j)e^{\sigma(\bm  z)}\sigma^i\partial_j a_i(\bm  z)\,.
\eeq
Then, substituting $\mathcal{G}\sim \mathcal{G}_0+\mathcal{G}_1+\ldots$ into~\eqref{eqgreen1} and using these expansions yields the following differential equation for  $\mathcal{G}_1$
\beq
\bigl(i\omega_n-H_0(z)\big)\mathcal{ G}_1( \bm  x,  \bm  x',\omega_n)=\bigl(\delta H(\bm  x)+e^{\sigma(\bm  z)}\delta \boldsymbol a(\bm  x)\cdot \boldsymbol\sigma\bigr)\,\mathcal{ G}_0( \bm  x,  \bm  x',\omega_n)\,.
\eeq
This equation is solved by the first term in a Schwinger-Dyson expansion
\beq\label{g1}
\mathcal{ G}_1( \bm  x,  \bm  x',\omega_n)=\int d^2 x''\, \mathcal{ G}_0( \bm  x,  \bm  x'',\omega_n)\bigl(\delta H(\bm  x'')+i\omega_n\,e^{\sigma(\bm  z)}\delta \boldsymbol a(\bm  x'')\cdot \boldsymbol\sigma\bigr)\mathcal{ G}_0( \bm  x'',  \bm  x',\omega_n)\,.
\eeq 
Noting that $\boldsymbol A=\boldsymbol a=0$ at the reference point $z$, we can use~\eqref{rel} to write
\beq
\mathcal{G}_0( \bm  x,  \bm  x',\omega_n)\Bigr|_{\boldsymbol A=\boldsymbol a=0}=\mathcal{D}_L(\bm  x)\Delta_0( \bm  x,  \bm  x',\omega_n)\,.
\eeq
We may also use the analogous relation
\beq
\mathcal{G}_0( \bm  x,  \bm  x',\omega_n)\Bigr|_{\boldsymbol A=\boldsymbol a=0}=\Delta_0( \bm  x,  \bm  x',\omega_n)\mathcal{D}_R(\bm  x')\,,
\eeq
where
\beq
\mathcal{D}_R(\bm  x')=(A_0(\bm  z)+i\omega_n)\mathbf{1}+m\, e^{\sigma(\bm  z)}\sigma_3+ie^{\sigma(\bm  z)}\sigma_i\frac{\overleftarrow{\partial}}{\partial x'_i}\,,
\eeq
to rewrite Eq.~\eqref{g1} in terms of the scalar Green function 
\beq\label{green1}
\begin{split}
&\mathcal{ G}_1( \bm  x,  \bm  x',\omega_n)=\\
\mathcal{D}_L(\bm  x)&\left[\int d^2 x''\, \Delta_0( \bm  x,  \bm  x'',\omega_n)\bigl(\delta H(\bm  x'')+i\omega_n\,e^{\sigma(\bm  z)}\delta \boldsymbol a(\bm  x'')\cdot \boldsymbol\sigma\bigr)\Delta_0( \bm  x'',  \bm  x',\omega_n)\right]\mathcal{D}_R(\bm  x')\,,
\end{split}
\eeq
with $\delta a_i(\bm x)=(x^j-z^j)\partial_j a_i(\bm  z)$.
 The integral over $x''$ in~\eqref{green1} can be turned into a Gaussian  by using the proper time representation of the scalar Green function
\beq
\Delta_0( \bm  x,  \bm  x',\omega_n)=-\frac{e^{-2\sigma}}{2\pi}K_0\left(b|\bm  x-\bm  x'|\right)=-\frac{e^{-2\sigma}}{2\pi}\int_0^\infty \frac{ds}{2s}e^{-\frac{| \bm  x-\bm  x'|^2}{4s}-b^2s}\,,
\eeq
where $b$ has been defined in Eq.~\eqref{b2}. The evaluation of $\mathcal{ G}_1( \bm  x,  \bm  x',\omega_n)$ is straightforward in principle but rather lengthy and is best carried out by computer.
 
According to Eq.~\eqref{ref} all the fields and their derivatives, including those present in $\mathcal{D}_L(\bm  x)$ and  $\mathcal{D}_R(\bm  x)$, are evaluated at the reference point $\bm z$, but  at  the end of the calculation we may set $ \bm  z\!=\! \bm  x$. We may also restore the whole dependence on $A_i$ and $a_i$, which have been set to zero at the reference point, by invoking gauge and KK invariance of  the components of the currents and energy-momentum tensor. In practice, this amounts to the substitution
\beq
\epsilon^{ij}\partial_i A_j\to\epsilon^{ij}(\partial_i A_j+a_i\partial_j  A_0)\,,
\eeq
which  is justified by noting that the right hand side  can be written as
\beq
\epsilon^{ij}(\partial_i A_j+a_i\partial_j  A_0)=\epsilon^{ij}\partial_i \tilde A_j+\epsilon^{ij}A_0\partial_i a_j\,,
\eeq
where $ \tilde A_i=A_i-A_0 a_i$. This is obviously KK and gauge invariant and reduces to the left hand side  for $a_i=0$. 
The complete expression for $\mathcal{ G}_1( \bm  x,  \bm  x',\omega_n)$ is rather cumbersome and  will not be given here. Instead, in the next section we will extract  the relevant parity odd pieces from $\mathcal{ G}_1( \bm  x,  \bm  x',\omega_n))$ and $\mathcal{ G}_0( \bm  x,  \bm  x',\omega_n)$.

\section{Computation of the  $U(1)$ current and  energy-momentum tensor}
\label{compcurr}
In this section  we will use the results  of the previous section  to derive the parity odd  contributions to the currents and  energy-momentum tensor. A look at Eq.~\eqref{green0} shows that there are no parity odd contributions to the Green function at leading order.
As a consequence,  in order to extract the parity-odd  contributions  to Eqs.~\eqref{curr} and~\eqref{tmunu} we must use the leading approximation $\mathcal G_0$ to the Green function for terms that contain explicit derivatives of the background fields and $\mathcal G_1$ otherwise. For instance, the formula for $T_{00}$ becomes
\beq
T_{00} = T_0 \sum_n \left[ e^\sigma (i \omega_n + A_0)\,  \text{tr}\, \mathcal{G}_1(\bm  x, \bm  x, \omega_n) - 
    \frac{1}{4} e^{3 \sigma}  \epsilon^{i j} \partial_i a_j\,   \text{tr}\, \sigma_3\, \mathcal{G}_0(\bm  x,\bm  x, \omega_n)  \right]\,.
\eeq

Then simple inspection of Eqs.~\eqref{curr} and~\eqref{tmunu} shows that only four traces of the Green functions  are needed. Using the results in the previous section for  $\mathcal{G}_0$ (Eq.~\eqref{green0}) and $\mathcal{G}_1$ (Eq.~\eqref{green1}), one finds that  the following combination  vanishes
\beq\label{deri}
\text{tr}\left( \frac{\partial}{\partial x^i}\mathcal{G}_1(\bm  x, \bm  x', \omega_n)  - 
         \frac{\partial}{\partial x^{\prime i}}\mathcal{G}_1(\bm  x, \bm  x', \omega_n) \right)  \biggl\lvert_{\bm  x' = \bm  x}=0\,,
\eeq
while the   remaining traces are given by 
\beq
\begin{split}
\text{tr}\, \mathcal{G}_1(\bm  x, \bm  x, \omega_n)&=-\frac{me^{-\sigma}}{4\pi}T_0 \sum_n\int_0^\infty ds\, e^{-b^2 s}\left[ 2\epsilon^{ij}(\partial_i A_j+a_i\partial_j  A_0)-(A_0+i\omega_n) \epsilon^{ij}\partial_i a_j\right]\,,\\
\mathrm{tr}\, \sigma_i\,\mathcal{G}_1(\bm  x, \bm  x, \omega_n)&=-\frac{me^{-2\sigma}}{2\pi}T_0 \sum_n\int_0^\infty ds\, e^{-b^2 s}\left[ \epsilon^{ij}\partial_j A_0-(A_0+i\omega_n) \epsilon^{ij}\partial_j \sigma\right]\,,\\
\mathrm{tr}\, \sigma_3\,\mathcal{G}_0(\bm  x, \bm  x, \omega_n)&=-\frac{me^{-\sigma}}{2\pi}T_0 \sum_n\int_0^\infty \frac{ds}{s} e^{-b^2 s}\,.
 \end{split}
 \eeq
As mentioned above, all the sum run over fermionic Matsubara frequencies $\omega_n=\frac{2\pi}{\beta} (n+\frac{1}{2})$ with  $\beta=1/T_0$. Note that these are the complete expressions for the traces and, as a consequence, all the contributions at first derivative order are parity violating. This is consistent with the fact that one can not construct a parity invariant   contribution to the partition function at this order.

Substitution of these expressions into Eqs.~\eqref{curr} and~\eqref{tmunu} gives the following formulas 
\beq\label{currents}
\begin{split}
J^i&= -\frac{me^{-2\sigma}}{2\pi}\epsilon^{ij}\left[I_0\partial_j A_0 -I_1\partial_j \sigma\right]\,,\\
  J_0&= \frac{m}{4\pi}\epsilon^{ij}\left[2I_0(\partial_i  A_j+a_i\partial_j  A_0 )  -I_1\partial_i a_j\right]\,,\\
 T_0^i&= \frac{me^{-2\sigma}}{4\pi}\epsilon^{ij}\left[I_1\partial_j A_0 -I_2\partial_j \sigma\right]\,,\\
  T_{00}&= -\frac{m}{4\pi}\epsilon^{ij}\left[2I_1(\partial_i  A_j+a_i\partial_j  A_0 )  -I_2\partial_i a_j\right]\,,\\
\end{split}
\eeq
where the first order contributions to the currents and energy-momentum tensor have been  expressed in terms of just three Matsubara sums
\beq\label{is}
\begin{split}
I_0&=T_0\sum_n\int_0^\infty ds e^{-b^2s}\\
I_1&=T_0\sum_n\int_0^\infty ds e^{-b^2s}(A_0+i\omega_n)\\
I_2&=T_0\sum_n\int_0^\infty \!\!ds e^{-b^2s}\!\left(\!\!(A_0+i\omega_n)^2\!\!+\!\frac{e^{2\sigma}}{2s}\right)\,,
\end{split}
\eeq
with $b^2= m^2-e^{-2\sigma}(A_0+i\omega_n)^2$ according to Eq.~\eqref{b2}.
 Doing the sum over Matsubara frequencies as described in the Appendix finally  yields the following results  
\begin{align}
 J^i&= \frac{e^{-\sigma}}{8\pi}f_-(\sigma,A_0)\epsilon^{ij}\partial_j A_0\ 
 -\frac{m}{8\pi} f_+(\sigma,A_0)\epsilon^{ij}\partial_j \sigma\label{ji}\,,\\
J_0&= -\frac{e^{\sigma}}{8\pi} f_-(\sigma,A_0)\epsilon^{ij}(\partial_i  A_j+a_i\partial_j  A_0 )+\frac{m e^{2\sigma} }{16\pi}f_+(\sigma,A_0)\epsilon^{ij} \partial_i a_j\label{j0}\,,\\
 T_0^i&=- \frac{m}{16\pi}f_+(\sigma,A_0)\epsilon^{ij}\partial_j A_0 +\frac{m^2 e^{\sigma} }{16\pi}f_-(\sigma,A_0)\epsilon^{ij}\partial_j \sigma +\frac{ m}{8\pi\beta}f_0(\sigma,A_0)\epsilon^{ij}\partial_j\sigma\label{t0i}\,,\\
  T_{00}&= \frac{me^{2\sigma}}{8\pi}f_+(\sigma,A_0)\epsilon^{ij}(\partial_i  A_j+a_i\partial_j  A_0 ) -\frac{m^2 e^{3\sigma} }{16\pi}f_-(\sigma,A_0)\epsilon^{ij}\partial_i a_j\\
  &\quad -\frac{m e^{2\sigma} }{8\pi\beta}f_0(\sigma,A_0)\epsilon^{ij}\partial_i a_j\label{t00}\,,
  \end{align}
where we have defined the functions
\beq\label{deff}
\begin{split}
f_\pm(\sigma,A_0)&= \tanh [\frac{\beta}{2}(A_0-e^\sigma m)]\pm\tanh[\frac{\beta}{2}(A_0+e^\sigma m)]\,,\\
f_0(\sigma,A_0)&=\log\left[2\cosh(A_0\beta)+2\cosh(e^\sigma\beta m)\right]\,.
\end{split}
\eeq

Equations~\eqref{ji} to \eqref{deff} are the main results in this section. It is worth mentioning that, as observed in the Appendix, the sums over Matsubara frequencies~\eqref{is} are finite without  subtractions, and so are all the 
contributions to the currents and energy-momentum tensor at first derivative order.

\section{The parity-odd equilibrium partition function}  
\label{partition}
In this section  our results for the parity odd contributions to the currents and energy-momentum tensor are used to obtain a completely explicit expression for the  equilibrium partition function. The general form of the parity odd partition function at first order in the derivative expansion has been given in~\cite{Banerjee:2012iz}
\beq\label{part}
\mathcal{W}=\frac{1}{2}\int d^2x \left({ \alpha}(\sigma,A_0)\epsilon^{ij}\partial_i \tilde A_j+T_0\beta(\sigma,A_0)\epsilon^{ij}\partial_i a_j\right)\,,
\eeq
where $ \tilde A_i=A_i-A_0 a_i$. Then using the variational formulae~\cite{Banerjee:2012iz}
\beq
\begin{split}
J^i=\frac{T_0}{\sqrt{-g}}\frac{\delta \mathcal{W}}{\delta\tilde A_i}\;\;\; &,\;\;\;
J_0=-\frac{e^{2\sigma}T_0}{\sqrt{-g}}\frac{\delta \mathcal{W}}{\delta A_0}\\
T_0^{\;i}=\frac{T_0}{\sqrt{-g}}\left(\frac{\delta \mathcal{W}}{\delta a_i}-\!\!A_0 \frac{\delta \mathcal{W}}{\delta\tilde A_i}\right)\;\;\; &,\;\;\; T_{00}=-\frac{e^{2\sigma}T_0}{\sqrt{-g}}\frac{\delta \mathcal{W}}{\delta \sigma}
\end{split}
\eeq
with  Eq.~\eqref{part} gives
\begin{align}
 J^i&= T_0\, e^{-\sigma}\epsilon^{ij}\partial_j\alpha\label{jia}\\ 
 J_0&=-T_0e^{\sigma}\left(\frac{\partial\alpha}{\partial A_0} \epsilon^{ij}\partial_i\tilde A_j+T_0
 \frac{\partial\beta}{\partial A_0} \epsilon^{ij}\partial_i  a_j\right)\\
 T_0^{\;i}&=T_0\, e^{-\sigma}\left( T_0\epsilon^{ij}\partial_j 
 \beta-A_0\epsilon^{ij}\partial_j 
 \alpha\right)\label{toiv}\\
 T_{00}&=-T_0e^{\sigma}\left(\frac{\partial\alpha}{\partial \sigma} \epsilon^{ij}\partial_i\tilde A_j+T_0
 \frac{\partial\beta}{\partial \sigma} \epsilon^{ij}\partial_i  a_j\right)\,.\label{tooa}
\end{align}
Using.~\eqref{ji} for $J_i$, Eq.~\eqref{jia} can be readily integrated, yielding
\beq\label{alpha}
\alpha(\sigma,A_0)=\frac{1}{4\pi}\log\left[ \cosh[\frac{\beta}{2}(A_0-e^\sigma m)]\right]-\frac{1}{4\pi}\log\left[ \cosh[\frac{\beta}{2}(A_0+e^\sigma m)]\right]\,.
\eeq
One can then use, for instance,  Eqs.~\eqref{t0i} and  \eqref{toiv} for $T_0^i$ to obtain
\beq\label{beta}
\begin{split}
\beta(\sigma,A_0)=&-\frac{\beta^2}{4\pi } A_0e^\sigma m+\frac{\beta}{8\pi}e^{\sigma}m\log[2\cosh(A_0\beta)+2\cosh(e^\sigma\beta m)]\\
&-\frac{\beta}{4\pi }(A_0+e^\sigma m)\log\left[1+e^{-\beta(A_0+e^\sigma m)}\right]\\
&+\frac{\beta}{4\pi }(A_0-e^\sigma m)\log\left[1+e^{-\beta(A_0-e^\sigma m)}\right]\\
&+\frac{1}{4\pi}\mathrm{Li}_2\left[-e^{-\beta(A_0+e^\sigma m)}\right]-\frac{1}{4\pi}\mathrm{Li}_2\left[-e^{-\beta(A_0-e^\sigma m)}\right]\,,
\end{split}
\eeq
where ${Li}_2$ is the polylogarithm function. It can be easily checked   that the expressions for $J_0$ and $T_{00}$ derived  from the partition function $\mathcal{W}$ agree with the results obtained in the last section. 
Note that the existence of a (unique) solution  implies non-trivial integrability conditions for the currents and energy-momentum tensor, thus providing a stringent check of the correctness of  our results.  Eqs.~\eqref{alpha} and \eqref{beta} are the main results in this paper.

In the limit of small fermion mass the functions ${ \alpha}$ and ${ \beta}$ take the following form 
\beq
\begin{split}
{ \alpha}(\sigma,A_0)&\to -\frac{m e^{\sigma}\beta}{4\pi} \tanh\left(\frac{A_0\beta}{2}\right)+\mathcal O(m^2)\\
{ \beta}(\sigma,A_0)&\to -\frac{m e^{\sigma}\beta}{8\pi}\left[-\log\left[2(1+\cosh(A_0\beta))\right]+2A_0\beta\tanh\left(\frac{A_0\beta}{2}\right) \right] +\mathcal O(m^2)
\end{split}
\eeq
and vanish for massless fermions. However, this does not imply the absence of parity violating effects in this limit~\cite{Redlich:1983kn}. The reason is that, even though the contributions to the currents and energy-momentum tensor at first derivative order are finite, there are divergent contributions at leading order. Specifically,  at zero derivative order $T_{00}$ is given by
\beq
T_{00}=T_0 \sum_n \left[ e^\sigma (i \omega_n + A_0)\,  \text{tr}\, \mathcal{G}_0(\bm  x, \bm  x, \omega_n)\right]=-\frac{e^{-\sigma}}{2\pi}T_0 \sum_n\int_0^\infty \frac{ds}{s}\, e^{-b^2 s}(A_0+i\omega_n)^2\,.
\eeq
The divergence can be extracted in the $T_0\to 0$ limit, where the sum over Matsubara frequencies reduces to an integral. In this limit we find
\beq\label{div}
 \left.T_{00}\right|_{T_0=0} =\lim_{\epsilon \to 0}\frac{e^{2\sigma}}{8\pi^{3/2}}\int_\epsilon^\infty  \frac{ds}{s^{5/2}}e^{-m^2 s}=\lim_{\epsilon \to 0}\frac{e^{2\sigma}}{12\pi^{3/2}}\left[\epsilon^{-3/2}-3 m^2\epsilon^{-1/2}+\ldots \right]\,,
\eeq
where the dots stand for contributions that are finite in the $\epsilon\to 0$ limit.

Thus the theory has to be regularized, and  this can be done in a gauge-invariant way  by means of  a Pauli-Villars regulator, which amounts to the introduction of auxiliary fermions with large masses $\{M_i\}$ and weights $\{C_i\}$. In order to cancel the divergences in Eq.~\eqref{div} we must impose
\beq\label{cond}
\sum_{i=0}^n C_i=0\;\;, \;\;\; \sum_{i=0}^n C_i M_i^2=0\,,
\eeq
where $i=0$ refers to the physical fermion, i.e., $C_0=1$ and $M_0=m$. A minimal system satisfying these constraints includes three auxiliary fermions with weights $C_1=C_2=-1$, $C_3=1$,  and large masses $M_1=M_2=M, M_3=\sqrt{2} M$ of the same sign as the mass of the physical fermion\footnote{Strictly speaking, Eq.~\eqref{cond} leaves the signs of the large masses undetermined. However, Pauli-Villars gives the correct coefficient for the radiatively induced Chern-Simons term in the effective action computed by other methods~\cite{Redlich:1983dv} only if the masses of the auxiliary fermions have the same sign as the physical fermion.}. Thus, in the regularized theory instead of Eq.~\eqref{part} we must consider
\beq\label{regpart}
\mathcal{W}^\text{reg}=\lim_{M\to\infty}\left(\mathcal{W}(m)-2\mathcal{W}(M)+\mathcal{W}(\sqrt{2} M)\right)\,.
\eeq
Now, the infinite mass limits 
\beq
\lim_{M\to\infty}{ \alpha}(\sigma,A_0)= -\frac{A_0\beta}{4\pi} \text{sgn}(M_i)\;,\;\;\;
\lim_{M\to\infty}{ \beta}(\sigma,A_0)= -\frac{A_0^2\beta^2}{8\pi} \text{sgn}(M_i)\,,
\eeq
where $\text{sgn}$ is the sign function, and  Eq.~\eqref{regpart} with $ \text{sgn}(M)= \text{sgn}(m)$, imply the following contributions from the auxiliary fields 
\beq
\delta{ \alpha}(\sigma,A_0)= \frac{A_0\beta}{4\pi} \text{sgn}(m)\;\;\;,\;\;\;
\delta{ \beta}(\sigma,A_0)= \frac{A_0^2\beta^2}{8\pi} \text{sgn}(m)\,.
\eeq
Then,  Eqs.~\eqref{jia}-\eqref{tooa} yield
\beq\label{pauli}
\delta J^i=-\frac{e^\sigma}{4\pi} \text{sgn}(m)\epsilon^{ij}\partial_jA_0\;\;\;,\;\;\;
\delta J_0=\frac{e^\sigma}{4\pi} \text{sgn}(m)\epsilon^{ij}(\partial_i  A_j+a_i\partial_j  A_0 )\,,
\eeq
together with $\delta T_{00}=\delta T_0^i=0$. Thus, at first order in the derivative expansion  we find  parity-odd contributions to the currents of massless fermions, but not to their energy-momentum tensor. Eq.~\eqref{pauli} differs from  the well known result for the currents in the absence of gravitational fields~\cite{Redlich:1983dv} by a term proportional to $\epsilon^{ij}a_i\partial_j  A_0$. This term represents a mixed gauge-gravitational contribution to the parity anomaly. As mentioned in the introduction, in $2+1$ dimensions the purely gravitational  contribution at zero temperature takes the form of the Cotton  tensor~\cite{AlvarezGaume:1984nf} which, being of third derivative order, can not be seen here. Note also that, for massive fermions,  
the contribution from the auxiliary fields~\eqref{pauli} must be added to the finite results in Eqs.~\eqref{ji}-\eqref{j0}.

\section{Discussion}
\label{discuss}
In this paper we have used perturbation theory combined with Schwinger's proper time method to obtain a derivative expansion for the thermal two-point Green function of an ideal gas of massive fermions in  non-trivial static gauge and gravitational  backgrounds in $2+1$ dimensions. After relating the currents and energy-momentum tensor to traces of the Green function~\eqref{curr}-\eqref{tmunu}, we have extracted all the parity violating contributions at first derivative order in the background fields. These contributions are finite without subtractions and are explicitly given by Eqs.~\eqref{ji}-\eqref{t00}. We have also shown that there are no parity preserving contributions at first order in the derivative expansion.

These results have been used to obtain a completely explicit expression for the  equilibrium partition function, determining the two unknown functions ${ \alpha}(\sigma,A_0)$ and ${ \beta}(\sigma,A_0)$ defined in~\cite{Banerjee:2012iz}. The role of a Pauli-Villars regulator has also been analyzed showing that, at  first derivative order,  
massless fermions have parity-odd contributions to their currents but not to their energy-momentum tensor~\eqref{pauli}.

As discussed in detail in~\cite{Banerjee:2012iz}, several adiabatic transport coefficients such as Hall electric and thermal conductivities can be derived form the  knowledge of the  functions ${ \alpha}(\sigma,A_0)$ and ${ \beta}(\sigma,A_0)$ in the equilibrium partition function~\eqref{part}. There are, however, additional adiabatic transport coefficients which can not be derived 
in this formulation. Among them is the Hall viscosity, which involves the spatial components $T^{ij}$ of the energy-momentum tensor and 
 has been connected with the intrinsic angular momentum density~\cite{Nicolis:2011ey} and  with the gravitational response  of the system  in the presence of geometric torsion~\cite{Hughes:2011hv,Hughes:2012vg}. For the torsionless backgrounds considered in this paper and in~\cite{Banerjee:2012iz}, the spatial components $T^{ij}$ vanish at first derivative order.
 
 Although the possible interplay between torsion and  Hall viscosity has been recently questioned~\cite{Haehl:2013kra}, it seems to us that a more detailed analysis  including non-zero torsion might be of interest. 
Indeed, in $2+1$ dimensions  the angular momentum density is proportional to $\overline{\psi} \psi$ and acquires a non-zero equilibrium expectation value, in contrast to the situation  in $3+1$ dimensions.   Since this  quantity is the zero component of a conserved current,  it should be possible to include it in the thermodynamic description by switching on the appropriate  conjugate  source, which must be a component of the contortion tensor (see, for instance, ref.~\cite{Hehl:1976kj}). This component  would play the role of  additional background data in the partition function. It is not clear to us that the spatial components of the energy-momentum tensor $T^{ij}$ would continue to vanish after these modifications. We hope to pursue these issues in future work.

\acknowledgments
This	work	is	supported in part by the Spanish Ministry of Science and Technology under Grant FPA2012-34456 and the Spanish Consolider-Ingenio 2010 Programme CPAN (CSD2007-00042), and by the Basque Government under Grant IT559-10.

\appendix

\section{Matsubara sums}
\label{matsu}
In this Appendix we evaluate the  three sums over the Matsubara frequencies $\omega_n=\frac{2\pi}{\beta} (n+\frac{1}{2})$ which appeared  in section~\ref{compcurr}:
\beq
\begin{split}
I_0&=T_0\sum_n\int_0^\infty ds e^{-b^2s}\\
I_1&=T_0\sum_n\int_0^\infty ds e^{-b^2s}(A_0+i\omega_n)\\
I_2&=T_0\sum_n\int_0^\infty \!\!ds e^{-b^2s}\!\left(\!\!(A_0+i\omega_n)^2\!\!+\!\frac{e^{2\sigma}}{2s}\right)\,,
\end{split}
\eeq
where $b^2= m^2-e^{-2\sigma}(A_0+i\omega_n)^2$. The evaluation of  $I_0$ and $I_1$ is straightforward if one does  the integrals first. For instance
\beq
\begin{split}
I_0=T_0\sum_n\int_0^\infty ds e^{-b^2s}&=T_0\sum_n \frac{1}{b^2}=T_0\sum_n \frac{1}{m^2-e^{-2 \sigma }
   (A_0+i\omega_n)^2}\\
&=   -\frac{e^\sigma}{4m} \left[ \tanh \frac{\beta}{2}(A_0-e^\sigma m)-\tanh\frac{\beta}{2}(A_0+e^\sigma m)\right] \,.
\end{split}   
\eeq
In the same way one obtains
\beq
I_1=-\frac{e^{2\sigma}}{4} \left[ \tanh \frac{\beta}{2}(A_0-e^\sigma m)+\tanh\frac{\beta}{2}(A_0+e^\sigma m)\right]\,.
\eeq
Obviously, we can not apply this method to $I_2$, as the integral of the second term diverges at the lower limit. Instead, we  use the identity
\beq
(A_0+i\omega_n)^2 e^{-b^2s}=e^{2\sigma}\left(\frac{d}{ds}+m^2\right)e^{-b^2s}
\eeq
to write
\beq\label{reg}
I_2=T_0\sum_n\int_0^\infty ds \left(\frac{d}{ds}+\frac{1}{2s}+m^2\right)e^{-b^2s+2\sigma}\,.
\eeq
The required sum evaluates to
\beq
T_0\sum_n e^{-b^2s+2\sigma}=\frac{e^{-m^2s+3\sigma}}{2\sqrt{\pi s}}\vartheta_3\left(\frac{1}{2}(\pi-iA_0\beta) ,e^{-\frac {e^{2\sigma}\beta^2}{4s}}\right)\,,
\eeq
where $\vartheta_3$ is a Jacobi $\Theta$ function, which admits the expansion
\beq\label{thetaser}
\vartheta_3(u,q)=1+2\sum_{n=1}^\infty q^{n^2}\cos(2nu)\,.
\eeq
This gives
\beq
 T_0\sum_n\left(\frac{d}{ds}+\frac{1}{2s}+m^2\right) e^{-b^2s+2\sigma}=
\frac{\beta^2e^{5\sigma}}{8\sqrt\pi s^{5/2}}\exp[{-\frac{e^{2\sigma}\beta^2}{4s}}]\vartheta'_3\left(\frac{1}{2}(\pi-iA_0\beta) ,e^{-\frac {e^{2\sigma}\beta^2}{4s}}\right)\,,
\eeq
where
\beq
\vartheta'_3(u,q)\equiv \frac{\partial\vartheta_3(u,q)}{\partial q}=2\sum_{n=1}^\infty n^2q^{n^2-1}\cos(2nu)\,.
\eeq
Now the integral over $s$ can be done term by term without encountering any divergence, giving
\beq
I_2=\sum_{n=1}^\infty e^{-e^\sigma |m| n \beta+3\sigma}\left(m+\frac{e^{-\sigma}}{n\beta}\right)(-)^n\cosh(A_0 n\beta)\,.
\eeq
Finally, summing the trigonometric series yields 
\beq
\begin{split}
I_2&=-\frac{me^{3\sigma}}{4}\left[ \tanh \frac{\beta}{2}(A_0-e^\sigma m)-\tanh\frac{\beta}{2}(A_0+e^\sigma m)\right]\\
&\quad-\frac{ e^{2\sigma} }{2\beta}\log\left[2\cosh(A_0\beta)+2\cosh(e^\sigma\beta m)\right]\,.
\end{split}
\eeq 
Note  that the three  sums computed in this Appendix are finite from the outset, without the need for infinite subtractions: the potential divergence in the integral of the second term of $I_2$ cancels against another divergence from the first term, and the net result is finite.

\bibliographystyle{JHEP}
\bibliography{bibgrav2+1}

\end{document}